\pdfoutput=1

\documentclass[11pt]{article}

\usepackage[]{acl}

\usepackage{times}
\usepackage{latexsym}

\usepackage[T1]{fontenc}

\usepackage[utf8]{inputenc}

\usepackage{microtype}
\usepackage{float}
\usepackage{graphicx}
\usepackage{cleveref}
\usepackage{multirow}
\usepackage{amssymb}
\usepackage{xcolor}
\usepackage{bbding}
\usepackage{pifont}
\definecolor{darkred}{rgb}{0.8, 0, 0}  
\definecolor{darkgreen}{rgb}{0, 0.5, 0} 
\usepackage{colortbl}
\usepackage{float}
\usepackage{array}
\newcolumntype{C}[1]{>{\centering\arraybackslash}m{#1}}

\usepackage{pifont}
\usepackage{xcolor}
\definecolor{ForestGreen}{RGB}{34,139,34}
\definecolor{BrickRed}{rgb}{.72,0,0}
\definecolor{LakeBlue}{RGB}{0,61,153}
\usepackage{makecell}
\usepackage{xspace}
\usepackage{booktabs}

\newcommand{\seek}{\textit{SeeClick}\xspace}
\newcommand{\seesp}{\textit{ScreenSpot}\xspace}

\usepackage[export]{adjustbox}
\usepackage{enumitem}

\title{\mbox{\protect\includegraphics[scale=.08, valign=c]{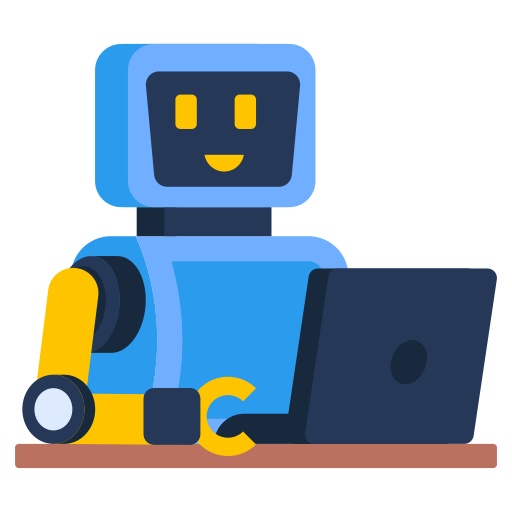} SeeClick: Harnessing GUI Grounding for Advanced Visual GUI Agents}}

\author{
Kanzhi Cheng\textsuperscript{$\diamondsuit$$\heartsuit$}\thanks{\, Work done during internship at Shanghai AI Laboratory.} \quad \quad
Qiushi Sun\textsuperscript{$\heartsuit$} \quad \quad
Yougang Chu\textsuperscript{$\diamondsuit$} \quad \quad
Fangzhi Xu\textsuperscript{$\heartsuit$} \\
\bf{
Yantao Li\textsuperscript{$\diamondsuit$} \quad \quad
Jianbing Zhang\textsuperscript{$\diamondsuit$} \quad \quad
Zhiyong Wu\textsuperscript{$\heartsuit$} 
}\\
\textsuperscript{$\diamondsuit$}Department of Computer Science and Technology, Nanjing University \\
\textsuperscript{$\heartsuit$}Shanghai AI Laboratory \\
\texttt{\{chengkz,chuyg,li\textunderscore yantao\}@smail.nju.edu.cn} \quad \texttt{qiushisun@u.nus.edu} \\
\texttt{fangzhixu98@gmail.com} \quad
\texttt{zjb@nju.edu.cn} \quad
\texttt{wuzhiyong@pjlab.org.cn} \\
}

\begin{document}
\maketitle
\begin{abstract}
Graphical User Interface (GUI) agents are designed to automate complex tasks on digital devices, such as smartphones and desktops. 
Most existing GUI agents interact with the environment through extracted structured data, 
which can be notably lengthy (e.g., HTML) and occasionally inaccessible (e.g., on desktops).  
To alleviate this issue, we propose a novel visual GUI agent -- \textit{SeeClick}, which only relies on screenshots for task automation.  
In our preliminary study, we have discovered a key challenge in developing visual GUI agents: GUI grounding -- the capacity to accurately locate screen elements based on instructions.
To tackle this challenge, we propose to enhance \textit{SeeClick} with GUI grounding pre-training and devise a method to automate the curation of GUI grounding data.
Along with the efforts above, we have also created \seesp, 
the first realistic GUI grounding benchmark that encompasses mobile, desktop, and web environments.
After pre-training, 
\textit{SeeClick} demonstrates significant improvement in \seesp over various baselines.
Moreover, comprehensive evaluations on three widely used benchmarks consistently support our finding that advancements in GUI grounding directly correlate with enhanced performance in downstream GUI agent tasks. \footnote{The model, data and code are available at \url{https://github.com/njucckevin/SeeClick}.}
\end{abstract}

\section{Introduction}
A perennial topic in machine intelligence is the development of Graphical User Interface (GUI) agent systems, 
like Siri and Copilot, to automate complex tasks on computing devices, thereby reducing human workload \citep{shi2017world, li2020mapping}.
Recent advances in Large Language Models (LLMs) such as GPT-4~\citep{openai2023gpt4}
have significantly propelled the evolution of GUI agents~\citep{gur2023real, zhou2023webarena}.
These agents interact with the environment by interpreting structured texts, e.g., HTML from webpages, then elicit LLM for planning, reasoning, and execution~\citep{kim2023language, zheng2023synapse}.


\begin{figure}[t]
\centering
\includegraphics[width=0.48\textwidth]{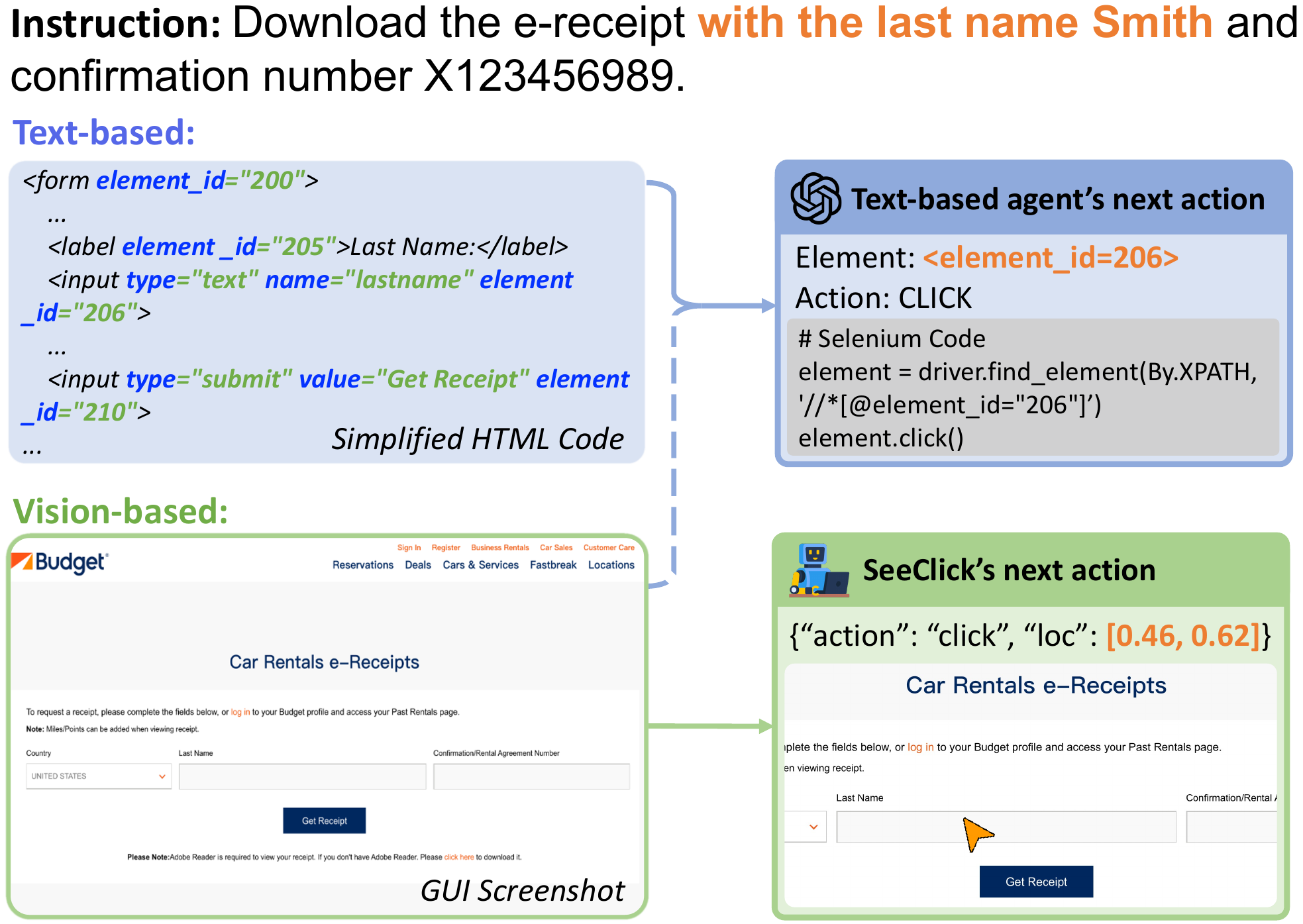}
\caption{Text-based agents select target elements from structured texts, occasionally augmented with screenshots. \textit{SeeClick} employs a vision-based methodology to predict action locations solely relying on screenshots.}
\label{fig:main_figure}
\end{figure}

However, GUI agents depend on structured text face three inherent limitations:
(1) Structured text is not always accessible, especially for iOS or desktop applications where acquiring such information is challenging~\citep{shaw2023pixels};
(2) The verbose nature of structured text 
constitutes an inefficient context for LLMs,
while also omitting crucial information such as layout, images, and icons~\citep{deng2023mind2web};
(3) The variety of structured text - including HTML, DOM, and Android VH - necessitates the curation of task-specific observation and action spaces~\citep{kim2023language, zhou2023webarena}.
These entrenched deficiencies in text-based approaches call for an alternative solution.

In this paper, 
we introduce \seek, 
a visual GUI agent built on Large Vision-Language Models (LVLMs).
Inspired by human interaction with GUIs, as illustrated in \Cref{fig:main_figure}, 
\seek is designed to perform low-level actions like clicking or typing directly by observing interface screenshots.
This innovative approach bypasses the interaction with cumbersome structured text, 
empowering \seek to universally adapt to various GUI platforms.
Building such visual agents presents a foundational challenge: GUI grounding - the capacity to accurately locate screen elements based on instructions, 
which is absent in current LVLMs.
To tackle this challenge, 
\seek enhances LVLM with a GUI grounding pre-training strategy. 
We devise a method to automate the curation of web grounding data and adapt public mobile UI datasets to obtain mobile grounding data.
\seek employs the above-curated dataset for continual pre-training of the LVLM, enabling it to accurately locate elements such as text, widgets, and icons in various GUI environments.

Given GUI grounding is a fundamental yet underexplored capacity for GUI agents, we establish \seesp, 
the first realistic GUI grounding evaluation benchmark across various GUI platforms.
\seesp contains over 600 screenshots and 1200 instructions from iOS, Android, macOS, Windows, and webpages, and specifically includes both text-based elements and a variety of widgets and icons.
Evaluation results confirm \seek's superiority over current LVLMs, validating the effectiveness of GUI grounding pre-training.

Finally, 
we adapt \seek to mobile and web agent tasks, including MiniWob~\citep{shi2017world}, AITW~\citep{rawles2023android}, and Mind2Web~\citep{deng2023mind2web}.
As a purely vision-based agent, \seek achieves impressive
performance. 
It surpasses the strong visual baseline Pix2Act
while utilizing merely 0.3\% training data.
Moreover, 
experimental results on these three benchmarks consistently support our findings that improvement in GUI grounding directly correlates with enhanced agent task performance.

Our main contributions are as follows:
\begin{itemize}[itemsep=2pt,topsep=3pt,parsep=0pt,leftmargin=*]
    \item We develop a unified visual GUI agent \seek, which solely relies on interface screenshots to perform clicking and typing actions across diverse GUI platforms.
    \item We prospectively explore GUI grounding for visual GUI agents, and enhanced \seek with proposed GUI grounding pre-training strategy.
    \item We create a realistic GUI grounding benchmark \seesp, encompassing more than 1200 instructions from various GUI platforms.
    \item Experimental results on \seesp and three agent tasks demonstrate that enhancing agents' grounding capacity is key to improving performance in downstream agent tasks.
\end{itemize}

\section{Related work}
\noindent
\textbf{Autonomous GUI Navigation}
Early research explored task automation in simplified web~\citep{shi2017world, liu2018reinforcement, gur2018learning} and mobile UI~\citep{li2020mapping, burns2022dataset, li2022spotlight}.
With LLM advancements~\citep[][\emph{inter alia}]{openai2023gpt4, touvron2023llama, xu2023symbol, sun2023corex, wu2024oscopilot},
LLM-centric agents have become the dominant paradigm.
A line of works focused on prompting
ChatGPT and GPT-4 for web tasks,
via in-context learning~\citep{zheng2023synapse} and self-refine~\citep{kim2023language}.
Other research explored training LLMs as specialized agents. 
\citet{deng2023mind2web} devised a two-stage method for identifying target elements within intricate HTML. 
\citet{gur2023real} proposed to interact with websites via programming.

Given the constraints of LLM to only process text,
recent efforts have attempted vision-based GUI navigation \citep{shaw2023pixels, zhan2023you, hong2023cogagent}.
These methods primarily utilize GPT-4V \citep{yan2023gpt, gao2023assistgui} and also require GUI metadata as input \citep{yang2023appagent, zheng2024gpt}. 
In this paper, we construct a universal visual GUI agent \seek by customizing open-sourced LVLM, capable of operating across various GUI platforms without needing any GUI metadata.

\noindent
\textbf{Large Vision-Language Models}
Recent research has invested tremendous effort in constructing LVLMs capable of jointly processing image and text \citep{liu2023llava, zhu2023minigpt, ye2023mplug, li2023otter},
integrating vision encoders with LLMs through connecting layers,
inheriting LLMs' linguistic and reasoning skills to perform vision-language tasks.
A series of studies focused on grounding with LVLMs \citep{wang2023visionllm, bai2023qwen, chen2023minigpt}, such as providing bounding boxes for objects when generating responses \citep{chen2023shikra, peng2023kosmos}.
Nonetheless, these efforts primarily addressed natural images and did not explore GUI contexts. 
This paper focuses on grounding in GUIs and explores the potential of LVLMs as visual agents.

\begin{figure*}[t!]
\centering
\includegraphics[width=\textwidth]{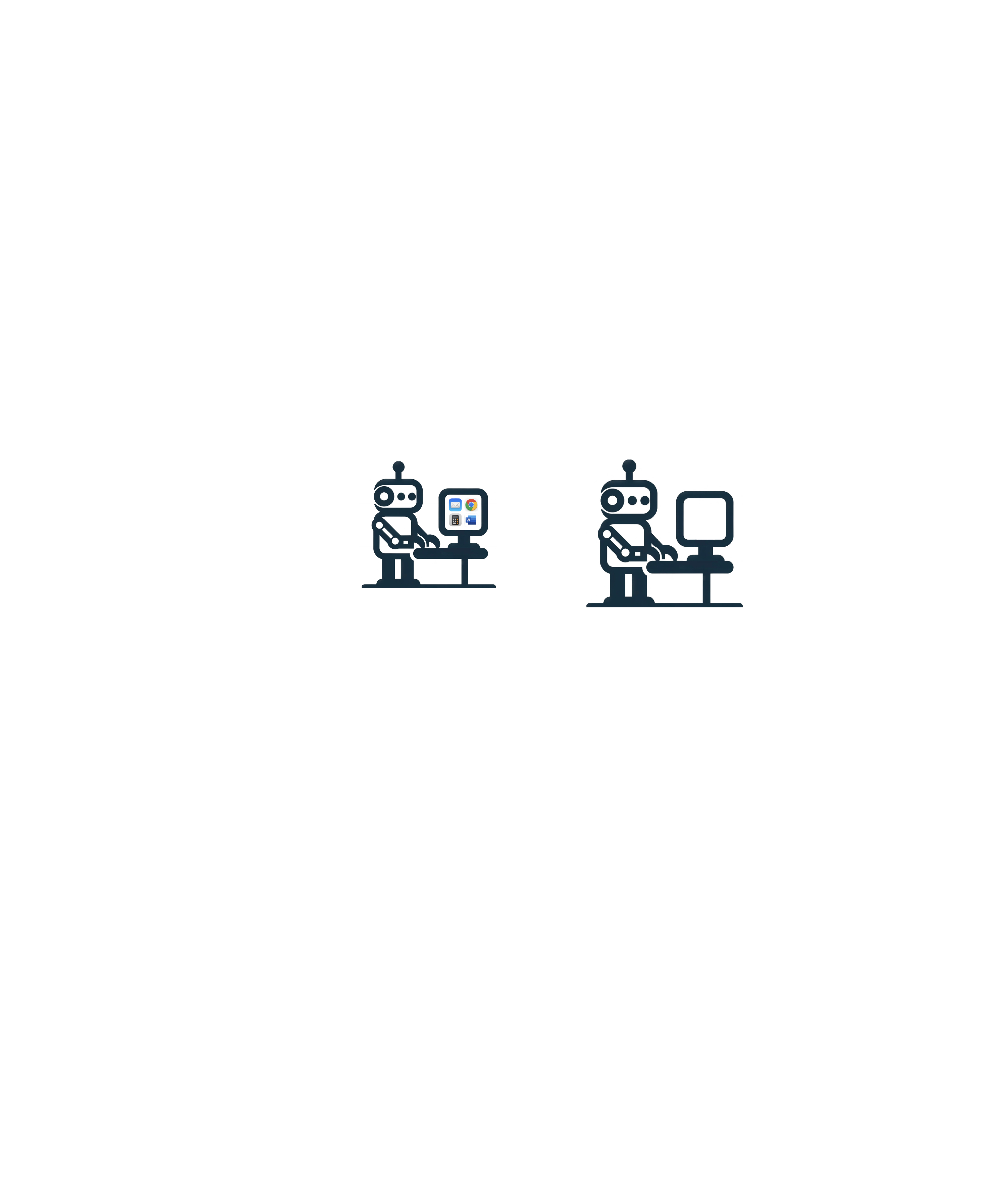}
\captionsetup{skip=3pt}
\caption{Overview of our universal visual GUI agent \seek. (a) depicts the framework of \seek and GUI grounding pre-training. (b) provides examples of \seesp across various GUIs and types of instructions. (c) displays the real-world application of \seek when adapted to downstream web agent tasks.}
\label{fig:framework}
\vspace{-1.0em}
\end{figure*}


\section{Approach}
\vspace{-0.3em}
Our preliminary study highlights a major challenge in developing visual GUI agents:
GUI grounding, the capacity to locate screen elements based on instructions.
Although recent LVLMs have claimed grounding capability on natural images \citep{bai2023qwen, wang2023visionllm}, 
GUI screenshots differ significantly with dense text and numerous icons and widgets.
These differences impair existing LVLMs' grounding performance in GUI contexts and 
limit their potential as visual GUI agents.

This paper seeks to harness LVLMs with GUI grounding skills, paving the way for a visual GUI agent that executes instructions only relying on screenshots.
As presented in \Cref{fig:framework}, \seek is a foundational model for GUIs, 
and tailored for adaption to agent tasks.
Next, we introduce the birth of \seek, including the formalization of GUI grounding task, the construction of continual pre-training data, and training details.

\subsection{GUI grounding for LVLMs}

As GUI grounding is the core capability of \seek, we first elucidate how to train LVLM for language generation to perform grounding tasks.
Given an interface screenshot $s$ and a collection of elements $\{(x_i, y_i)|_i\}$ on it, where $x_i$ denotes the textual description of the $i$-th element and $y_i$ indicates the element's location (represented as a bounding box or point).
As depicted in \Cref{fig:framework}(a), LVLM predicts the location of the element $y$ based on the interface screenshot $s$ and its textual description $x$, i.e. calculating $p(y|s,x)$.

A potential challenge is how LVLMs predict numerical coordinates in a language generation format. 
Previous studies~\citep{chen2021pix2seq, wang2023visionllm, shaw2023pixels} divide the image into 1000 bins, and creating a new 1,000-token vocabulary $\{<p0>, <p1>, ..., <p999>\}$ to represent the x and y coordinates.
In this work, we adopt a more intuitive manner used in LVLMs \citep{chen2023shikra, bai2023qwen}, treating numerical values as natural languages without any additional tokenization or pre-/post-processing.
For instance, in \Cref{fig:framework}(a), for a smartphone screenshot and the instruction ``View the new album of Jony J'', we craft a query prompt: ``In the UI, where should I click if I want to <instruction>?''. 
Subsequently, we normally compute the cross-entropy loss between the model output and the ground truth ``click (0.49, 0.40)'' to optimize the LVLM.

\vspace{-0.30em}
\subsection{Data Construction}
\vspace{-0.10em}
\label{sec:data}
We train \seek using three collections of data: web UI data crawled from the internet, mobile UI data reorganized from public datasets and general vision-language instruction-following data.

\begin{figure}[t]
\centering
\includegraphics[width=0.40\textwidth]{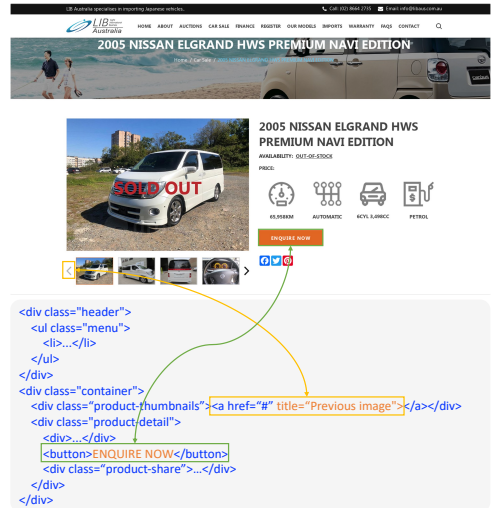}
\captionsetup{skip=2pt}
\caption{Example of two types of elements automatically collected from the webpage.}
\label{fig:webdata}
\vspace{-0.8em}
\end{figure}

\noindent
\textbf{Web Data.} Web UIs, featuring a variety of layouts and design styles across websites, are ideal for training LVLMs' general recognition and grounding capabilities across different GUI contexts.
We collect approximately 300k web pages from the latest Common Crawl repository
to serve as our training data for web UI.
For each webpage $s$, we collect two types of elements from the HTML code as exemplified in \Cref{fig:webdata}: 
(1) elements that display visible text content; 
and (2) elements with a special ``title'' attribute that display descriptive text when hovering. This method ensures that we gather a series of interactable elements $y$ and their corresponding instructions $x$, while encompassing a wide range of text and icon elements.
In addition to the grounding task $p(y|s,x)$, we also include web OCR task $p(x|s,y)$, predicting text description based on coordinates.

\noindent
\textbf{Mobile Data.} For mobile UI, we include three types of data: widget captioning, mobile UI grounding, and mobile UI summarization.
The widget captioning dataset provides language descriptions for mobile UI elements; for example, the description ``play music'' for the play button on a music player interface. 
We utilize the training split of the dataset provided by \citep{li2020widget}, 
containing nearly 20k screenshots, 40k widgets, and 100k descriptions.
We derive mobile UI grounding data by reversing the process of widget captioning, 
treating language descriptions as instructions and corresponding widgets as target elements. 
To improve diversity, we also incorporate the automatically collected elements and instructions from RICO~\citep{li2020mapping}.
The mobile data involves diverse elements and instructions, facilitating the generalization of \seek's grounding proficiency to diverse GUI contexts.
We finally include mobile UI summarization data \citep{wang2021screen2words} to enhance overall interface comprehension.

\noindent
\textbf{General Data.} To maintain LVLM's general capacities on natural images, we incorporate the general vision-language instruction-following data from LLaVA~\citep{liu2023llava}, covering conversation, detailed description, and complex reasoning.

Finally, we mix the data above and craft 30 task-specific prompts for each added GUI task, 
resulting in a 1M dataset to train \seek.

\vspace{-0.30em}
\subsection{Training Details}
\vspace{-0.10em}
We build \seek through continual pre-training on a recent advanced LVLM, Qwen-VL \citep{bai2023qwen}, which possesses grounding capabilities and a higher resolution of 448*448.
We train Qwen-VL on the dataset we constructed (as described in \Cref{sec:data}) for about 10k steps (around 1 epoch) to obtain our GUI base model \seek.
During training, we employ LoRA~\citep{hu2021lora} to fine-tune both the visual encoder and LLM.
Further details and task examples are provided in \Cref{sec:pretraining}.


\begin{table*}[t!]
\centering
\renewcommand\arraystretch{1.1}
\tabcolsep=0.05cm
{\fontsize{10pt}{12pt}\selectfont
\begin{tabular}{cccccccccc}
\hline
         \multirow{2}{*}{LVLMs} & \multirow{2}{*}{\parbox{1.5cm}{\centering Model\\Size}}   & \multirow{2}{*}{\parbox{1.5cm}{\centering GUI\\Specific}} & \multicolumn{2}{c}{Mobile}        & \multicolumn{2}{c}{Desktop}       & \multicolumn{2}{c}{Web}           & \multirow{2}{*}{Average} \\ \cline{4-9}
         &    &                               & Text            & Icon/Widget     & Text            & Icon/Widget     & Text            & Icon/Widget     &                          \\ 
\hline
MiniGPT-v2     & 7B   & \textcolor{darkred}{\ding{55}}                              &  8.4\%               & 6.6\%                & 6.2\%                & 2.9\%                & 6.5\%                & 3.4\%                & 5.7\%                         \\
Qwen-VL  & 9.6B   & \textcolor{darkred}{\ding{55}}                             & 9.5\%           & 4.8\%           & 5.7\%           & 5.0\%           & 3.5\%           & 2.4\%           & 5.2\%                    \\ 
GPT-4V   & -    & \textcolor{darkred}{\ding{55}}     & 22.6\%    & 24.5\%    & 20.2\%    & 11.8\%    &9.2\%  & 8.8\%     & 16.2\% \\
\hline
Fuyu     & 8B   & \textcolor{darkgreen}{\ding{51}}                               & 41.0\%          & 1.3\%           & 33.0\%          & 3.6\%           & 33.9\%          & 4.4\%           & 19.5\%                   \\
CogAgent & 18B   & \textcolor{darkgreen}{\ding{51}}                               & 67.0\%          & 24.0\%          & \textbf{74.2\%} & 20.0\%          & \textbf{70.4\%} & 28.6\% & 47.4\%                   \\
\seek & 9.6B   & \textcolor{darkgreen}{\ding{51}}                              & \textbf{78.0\%} & \textbf{52.0\%} & 72.2\%          & \textbf{30.0\%} & 55.7\%          & \textbf{32.5\%} & \textbf{53.4\%}          \\ 
\hline
\end{tabular}
}
\caption{Results of different LVLMs on \seesp. The best results in each column are highlighted in \textbf{bold}. Benefiting from efficient GUI grounding pre-training, \seek significantly enhanced LVLMs' ability to locate GUI elements following instructions, and surpassed the strong baseline CogAgent with a smaller model size.}
\label{tab:screenspot}
\vspace{-1.0em}
\end{table*}


\section{\seesp: A Grounding Benchmark}
\vspace{-0.3em}

We recognize GUI grounding proficiency as essential for constructing visual GUI agents.
However, the constrained capabilities of earlier vision-language models resulted in limited attention, with scant research \citep{li2021vut, li2022spotlight, zhang2023reinforced} largely confined to an Android dataset \citep{deka2017rico} collected in 2017.

\begin{figure}[t]
\centering
\includegraphics[width=0.48\textwidth]{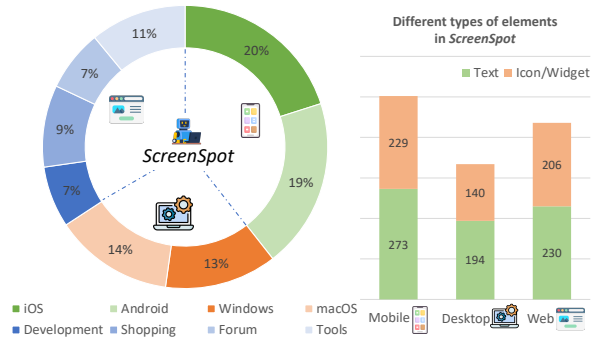}
\captionsetup{skip=3pt}
\caption{Statistic of our proposed GUI grounding benchmark \seesp. The left illustrates the diverse GUI environments included. The right displays the types of elements within each GUI category.}
\label{fig:screenspot}
\vspace{-0.8em}
\end{figure}

To address this research gap, we introduce \seesp, an up-to-date, realistic grounding evaluation benchmark encompassing various GUI platforms. 
It is designed to assess vision-language models' ability to locate screen elements based on instructions (\Cref{fig:framework}(b) provides some examples). 
\seesp has two distinctive features: (1) Various GUI platforms. It includes over 600 interface screenshots from mobile (iOS, Android), desktop (macOS, Windows), and web platforms, along with 1200+ instructions and corresponding actionable elements; (2) Icons/Widgets. \seesp includes a substantial number of icons and widgets in each GUI, which is more challenging to locate than texts (statistics are in \Cref{fig:screenspot}).
See \Cref{app:screenspot} for annotation details and examples.

To measure models' effectiveness in real-world scenarios, 
\seesp is carefully curated to ensure the samples are novel and not included in existing training resources.
We recruited experienced annotators to collect GUI interfaces and label instructions along with the bounding boxes for actionable elements.
For mobile and desktop, annotators were asked to select commonly used apps and operations; for web, we chose several types of websites  (development, shopping, forum, and tools) from the web environment WebArena~\citep{zhou2023webarena}.

\section{Experiments}

In this section, 
we first evaluate the GUI grounding capabilities of representative LVLMs and our proposed \seek.
Subsequently, we adapt \seek to mobile and web agent tasks, analyzing the correlation between the advanced grounding capacity and downstream task performance, while exploring the potential of purely vision-based GUI agents.

\subsection{GUI Grounding on \seesp}
As the foundation of visual GUI agents, GUI grounding has not received adequate attention in current LVLMs evaluations \citep{liu2023mmbench, yu2023mm}. Therefore, we evaluate LVLMs on our GUI-specific benchmark \seesp.

\noindent
\textbf{Compared LVLMs \& Evaluation.} We primarily evaluated two types of LVLMs: (1) Generalist LVLMs capable of tasks such as dialogue, recognition and grounding, including MiniGPT-v2 \citep{chen2023minigpt}, Qwen-VL \citep{bai2023qwen} and GPT-4V; (2) Recently released LVLMs specifically designed for GUI tasks, including Fuyu \citep{fuyu-8b} and CogAgent \citep{hong2023cogagent}.

Considering that GUI agents require clicking on the correct position, we calculate the click accuracy as the metric, defined as the proportion of test samples where the model predicted location falls in the ground truth element bounding box \citep{li2022grounded, zhang2023reinforced}. 
More details about evaluation on \seesp is in \Cref{app:screenspot}.

\noindent
\textbf{Results.} As shown in \Cref{tab:screenspot}, while generalist LVLMs have excelled in natural image grounding, 
their GUI grounding performance on \seesp is poor due to the significant differences between GUIs and natural images.
Even GPT-4V struggles with accurately locating screen elements.

In comparison, GUI-specific LVLMs have significant improvements. 
\seek achieved the best average performances across GUI platforms and two types of elements, even with fewer parameters than CogAgent.
This demonstrates the efficiency of our GUI grounding pre-training; with the rich UI elements and diverse instructions collected from the web and mobile, \seek quickly learns to understand human instructions for element localization, even in completely unseen scenarios like iOS and desktop.
\seek exhibits slightly inferior performance in locating text within desktop and web compared to CogAgent, possibly due to lower resolution and much smaller training data.
Notably, all models struggle with locating icons/widgets, highlighting the difficulty of identifying and grounding non-text elements on GUIs, which is the unique challenge posed by \seesp.


\begin{figure*}[t!]
\centering
\includegraphics[width=\textwidth]{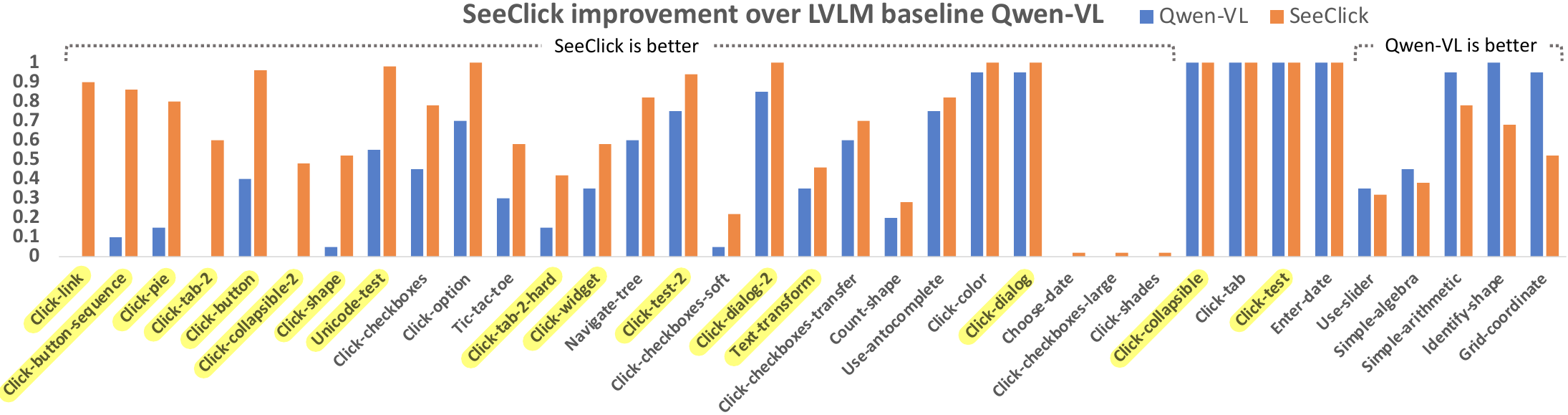}
\captionsetup{skip=1.5pt}
\caption{Comparison of \seek and Qwen-VL on MiniWob. 
Tasks marked with yellow shadows feature dynamic webpage layouts, simulating real-world GUI agent applications (details in appendix \Cref{miniwob_case}). \seek outperformed Qwen-VL in most tasks, highlighting the effectiveness of GUI grounding pre-training.}
\label{fig:seeclick_vs_qwen}
\end{figure*}

\vspace{-0.3em}
\subsection{Visual GUI Agent Tasks}
\vspace{-0.2em}
\label{sec:agent_task}
This section explores \seek's application to mobile and computer agent tasks: MiniWob, AITW, and Mind2Web.
We trained \seek on the respective training splits and tested it on the test sets.
Across these tasks, with provided instructions and memory of previous actions, \seek determines the next action solely by observing interface screenshots. 
The detailed task settings, action formats and interaction examples are in \Cref{app:agent_tasks}.

\vspace{-0.3em}
\subsubsection{MiniWob}
MiniWob \cite{shi2017world} comprises about 100 types of web automation tasks, where the agent is asked to interact with a simplified web environment to accomplish human instructions.
Existing open-source training data often lacks corresponding interface screenshots \citep{furuta2023multimodal}. Therefore, we rollout 50 successful episodes using an LLM strategy for each task in \citep{zheng2023synapse}, resulting in a 2.8K episodes dataset to train \seek.

\noindent
\textbf{Compared Methods \& Evaluation.} We compared \seek with a range of offline training methods.
Among these, the state-of-the-art method WebGUM~\citep{furuta2023multimodal} uses screenshots as auxiliary input but still selects HTML elements as actions.
Pix2Act~\citep{shaw2023pixels} is the only prior vision-based approach, trained with extensive demonstration data to perform actions.
To verify the effectiveness of GUI grounding pre-training, we also report the results of the LVLM baseline Qwen-VL when trained with the same 2.8K dataset.

Due to the variance in evaluation task sets among different methods \citep{liu2018reinforcement, furuta2023multimodal, shaw2023pixels}, for fairness, we report performance in two groups based on the overlapping MiniWob tasks.
We compute the success rate over 50 random seeds for each task and then compute the mean over all tasks as the final score. We provided task-wise scores in \Cref{app:miniwob_tasks}.

\begin{table}[t!]
\centering
\renewcommand\arraystretch{1.2}
\tabcolsep=0.06cm
{\fontsize{10pt}{12pt}\selectfont
\begin{tabular}{cccc}
\hline
Methods                 & Modality   & Dataset & Score          \\ \hline
\rowcolor{gray!25}
\multicolumn{4}{c}{Compared with text-based models over 45 tasks}      \\ \hline
CC-Net (SL)             & DOM+Image  & 2.4M    & 35.6\%                \\
WebN-T5                 & HTML       & 12K     & 55.2\%                \\
MM-WebN-T5              & HTML+Image & 347K    & 63.4\%                \\
WebGUM                  & HTML+Image & 2.8K    & 65.5\%                \\
WebGUM                  & HTML+Image & 347K    & \textbf{86.1\%}       \\
\seek                & Image      & 2.8K    & {\underline{73.6\%}}          \\ \hline
\rowcolor{gray!25}
\multicolumn{4}{c}{Compared with vision-based models over 35 tasks}    \\ \hline
CC-Net (SL)             & Image      & 2.4M    & 23.4\%                \\
Pix2Act                 & Image      & 1.3M    & 64.6\%                \\
Qwen-VL                 & Image      & 2.8K    & 48.4\%                \\
\seek                & Image      & 2.8K    & {\underline{\textbf{67.0\%}}} \\ \hline
\end{tabular}
}

\vspace{-0.5em}
\caption{Average scores of different methods on MiniWob. The best results in each setting are \textbf{bold}. Methods achieving the highest performance with limited data are \underline{underlined}. \seek outperforms a range of offline training methods as a purely vision-based model.}
\label{tab:miniwob}
\vspace{-1.2em}
\end{table}

\begin{table*}[t]
\centering
\renewcommand\arraystretch{1.1}
\tabcolsep=0.10cm
{\fontsize{10pt}{12pt}\selectfont
\begin{tabular}{ccccccccc}
\hline
Methods     & Modality & General       & Install       & GoogleApps    & Single        & WebShopping   & Overall       & ClickAcc     \\ \hline
ChatGPT-CoT & Text         & 5.9           & 4.4           & 10.5          & 9.4           & 8.4           & 7.7           & -             \\
PaLM2-CoT  & Text         & -             & -             & -             & -             & -             & 39.6          & -             \\
GPT-4V      & Image         & 41.7          & 42.6          & 49.8          & \textbf{72.8} & 45.7          & 50.5          & -             \\ \hline
Qwen-VL     & Image         & 49.5          & 59.9          & 46.9          & 64.7          & 50.7          & 54.3          & 57.4          \\
\seek    & Image         & \textbf{54.0} & \textbf{66.4} & \textbf{54.9} & 63.5          & \textbf{57.6} & \textbf{59.3} & \textbf{66.4} \\ \hline
\end{tabular}
}
\caption{Average scores of different methods on AITW. ClickAcc calculates the accuracy of click operation. The best results in each column are \textbf{bold}. \seek exhibits the best performance among competing baselines.}
\vspace{-0.75em}
\label{tab:aitw}
\end{table*}

\noindent
\textbf{Results.} As depicted in \Cref{tab:miniwob}, purely vision-based \seek surpassed strong baselines with substantially less training data.
Notably, with an equivalent amount of 2.8K training data, it outperformed the offline sota WebGUM, which uses both HTML and screenshots as input.
Moreover, thanks to LVLM's powerful reasoning and planning abilities and our GUI grounding pre-training, \seek exceeded the sota visual method Pix2Act, using less than $0.3\%$ training data.

Furthermore, \seek significantly surpassed the LVLM baseline Qwen-VL by nearly 20 percentage points, underscoring the importance of GUI grounding in boosting LVLM's performance.
To analyze in detail, we provide task-level comparisons in \Cref{fig:seeclick_vs_qwen}.
\seek notably excelled in tasks with dynamic interface layouts and element positions, confirming our hypothesis that general LVLMs struggle with accurately clicking, and \seek markedly improves this aspect.

\subsubsection{AITW}
\label{sec:aitw_split}
We evaluate \seek in smartphone environments with Android automation dataset Android In The Wild (AITW) \citep{rawles2023android}, which encompasses 30k instructions and corresponding 715k operation trajectories.
Previous approaches split train/val/test episode-wise, which poses a clear risk of overfitting due to: (1) instructions in the test set have appeared in training, and (2) an average of 20 similar trajectories per instruction.
In this work, we opt for an instruction-wise split, with 545/688/306/700/700 instructions from General/Install/GoogleApps/Single/WebShopping respectively, and retain one trajectory per instruction.
We selected 80\% for training and the remaining for testing in each subset.
This split avoids overfitting and reflects the performance of agents on unseen instructions. 
Further details and results on the original split are in \Cref{sec:aitw_origin}.

\noindent
\textbf{Compared Methods \& Evaluation.} We compare \seek with two types of baselines: (1) API-based LLMs 
such as ChatGPT-CoT \citep{zhan2023you}, PaLM2-CoT \citep{rawles2023android} and the latest GPT-4V \citep{yan2023gpt}; 
(2) Our trained LVLM baseline Qwen-VL.

We follow \citet{rawles2023android} to adopt the screen-wise action matching score as the main metric and additionally compute the click accuracy (ClickAcc), which calculates the accuracy when both reference and prediction are click operations.

\begin{table*}[t!]
\centering
\renewcommand\arraystretch{1.2}
\tabcolsep=0.10cm
{\fontsize{10pt}{12pt}\selectfont
\begin{tabular}{ccccccccccc}
\hline
\multirow{2}{*}{Methods} & \multirow{2}{*}{w/o HTML} & \multicolumn{3}{c}{Cross-Task}                & \multicolumn{3}{c}{Cross-Website}             & \multicolumn{3}{c}{Cross-Domain}              \\ \cline{3-11} 
                         &                           & Ele.Acc       & Op.F1         & Step SR       & Ele.Acc       & Op.F1         & Step SR       & Ele.Acc       & Op.F1         & Step SR       \\ \hline
MindAct (gen)            & \textcolor{darkred}{\ding{55}}                          & 20.2          & 52.0          & 17.5          & 13.9          & 44.7          & 11.0          & 14.2          & 44.7          & 11.9          \\
MindAct                  & \textcolor{darkred}{\ding{55}}                          & \textbf{55.1} & 75.7          & \textbf{52.0} & \textbf{42.0} & 65.2          & \textbf{38.9} & \textbf{42.1} & 66.5          & \textbf{39.6} \\
GPT-3.5-Turbo                  & \textcolor{darkred}{\ding{55}}                          & 20.3          & 56.6          & 17.4          & 19.3          & 48.8          & 16.2          & 21.6          & 52.8          & 18.6          \\
GPT-4                    & \textcolor{darkred}{\ding{55}}                          & 41.6          & 60.6          & 36.2          & 35.8          & 51.1          & 30.1          & 37.1          & 46.5          & 26.4          \\ \hline
Qwen-VL                  & \textcolor{darkgreen}{\ding{51}}                          & 15.9          & 86.7 & 13.3          & 13.2          & \textbf{83.5} & 9.2          & 14.1           & 84.3 & 12.0           \\
\seek                 & \textcolor{darkgreen}{\ding{51}}                          & \underline{28.3}          & \textbf{87.0} & \underline{25.5}          & \underline{21.4}          & 80.6          & \underline{16.4}          & \underline{23.2}          & \textbf{84.8}                & \underline{20.8}  \\ \hline
\end{tabular}
}
\caption{Comparsion of methods on Mind2Web. The best results in each column are \textbf{bold}. Improvements of \seek over LVLM baseline are \underline{underline}, with GUI grounding pre-training nearly doubling the step success rate.}
\vspace{-0.75em}
\label{tab:mind2web}
\end{table*}

\noindent
\textbf{Results.} As illustrated in \Cref{tab:aitw}, \seek achieved the best average performance among both API-based LLMs and trained LVLMs.
Specifically, \seek exhibited a 9\% increase in click accuracy over Qwen-VL, supporting the idea that GUI grounding enhances agent task performance through precise clicking.

\vspace{-0.30em}
\subsubsection{Mind2Web}
To assess \seek's capabilities in web navigation, we utilize the recently introduced Mini2Web dataset \citep{deng2023mind2web},
which comprises over 2000 open-ended tasks collected from 137 real websites, each with high-level instruction and corresponding human action trajectory.
Mind2Web was originally designed for text-based agents, which select actionable elements from simplified HTML in each step.
This work explores visual web agents that predict click positions directly from screenshots. 
For this purpose, we parsed screenshots and target element bounding boxes from the raw dump of Mind2Web.
To the best of our knowledge, this is the first attempt of web agents relying solely on screenshots as inputs for navigating real websites.

\noindent
\textbf{Compared Methods \& Evaluation.} We compare with html-based web agents Mind2Act~\citep{deng2023mind2web} and our visual baseline Qwen-VL.
Mind2Act employs a two-stage method, where a small LM first generates candidate elements from raw HTML, then a large LM selects the target via multi-choice QA; Mind2Act (gen) directly generates the target element instead.
GPT-3.5 and GPT-4 adopt the same multiple-choice QA formulation and include three demonstrations for in-context learning.

We calculate element accuracy (Ele.Acc), Operation F1 (Op.F1) and step success rate (Step SR). For vision-based methods, a prediction is considered correct if the predicted coordinate falls in the target element's bounding box. All other settings are following~\citep{deng2023mind2web}.

\noindent
\textbf{Results.} As displayed in \Cref{tab:mind2web}, \seek nearly doubled the Ele.Acc and Step SR compared to Qwen-VL.
This indicates that \seek's improvement in GUI grounding correlates with enhanced performance in web agent tasks.
HTML-based methods yield lower Op.F1 as around 20\% of groundturth elements are filtered out during candidate generation.
Although \seek can operate without extra HTML information, its performance trails sota HTML-based methods, since predicting click coordinates is much more difficult than choosing from HTML candidates.
This highlights the difficulty of grounding in intricate interfaces, suggesting substantial room for improvement in visual agents for real-world application.

\vspace{-0.30em}
\subsubsection{Grounding and Agent Performance}
\begin{figure}[t!]
\centering
\includegraphics[width=0.48\textwidth]{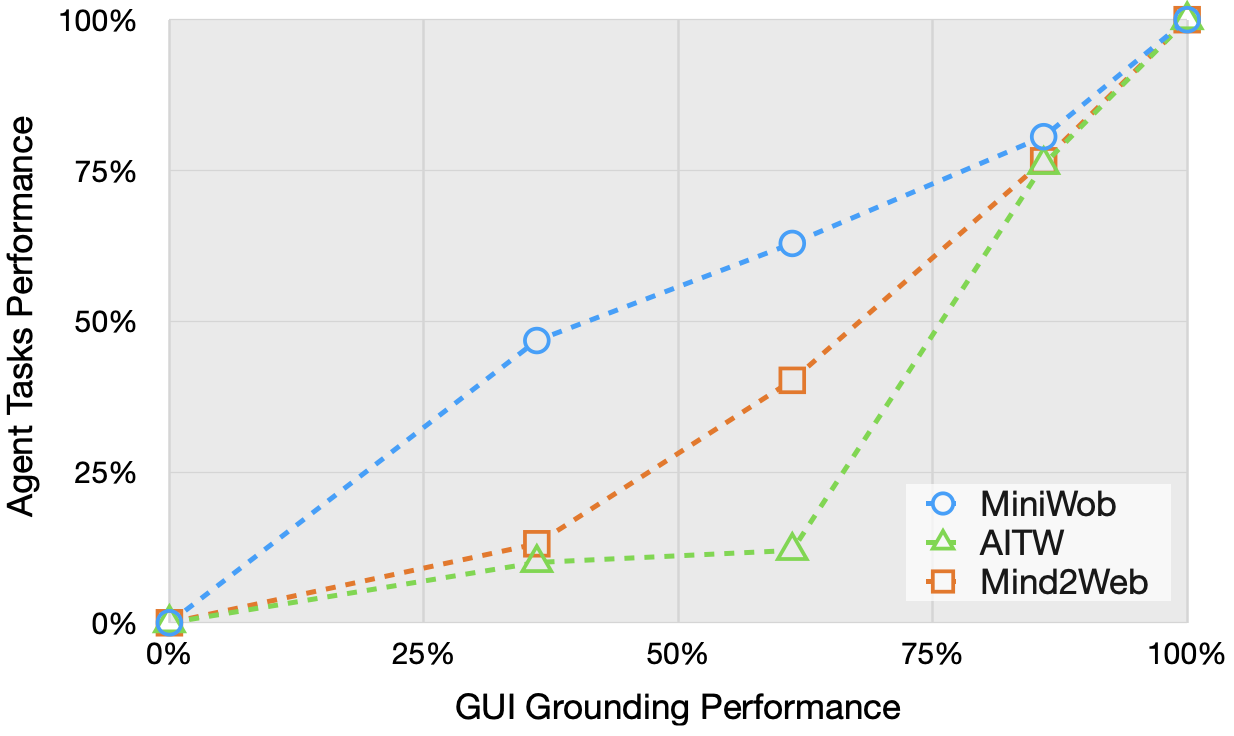}
\captionsetup{skip=2pt}
\caption{The correlation between agent tasks performance improvement and enhanced grounding ability.}
\label{fig:grounding2agent}
\end{figure}

To investigate the correlation between grounding and agent performance, we analyze the average score improvements of several \seek's checkpoints on \seesp and three downstream tasks. As depicted in \Cref{fig:grounding2agent},
enhanced GUI grounding capacity consistently boosts agent task performance, highlighting its crucial role in developing advanced visual GUI agents.

\subsubsection{\seek as Unified GUI Agent}
To access the potential of vision-based solutions in unifying GUI agent tasks, we evaluated jointly training \seek on three downstream tasks.
As shown in \Cref{tab:unified}, the unified model exhibited a slight performance decline, possibly due to the significant distinct interface of different GUIs.

\begin{table}[t!]
\centering
\renewcommand\arraystretch{1.2}
\tabcolsep=0.06cm
{\fontsize{10pt}{12pt}\selectfont
\begin{tabular}{lccc}
\hline
                   & MiniWob       & AITW          & Mind2web      \\ \hline
Qwen-VL\textsubscript{\textit{separate}} & 48.4          & 54.3          & 11.5          \\
\textit{SeeClick}\textsubscript{\textit{separate}} & 67.0          & 59.3          & 20.9          \\
\textit{SeeClick}\textsubscript{\textit{unified}}  & 64.1 & 57.1 & 19.5 \\ \hline
\end{tabular}
}
\caption{Separate v.s. unified training performance.}
\vspace{-1.0em}
\label{tab:unified}
\end{table}

\section{Conclusion}
In this paper, we introduce a visual GUI agent - \seek, which only relies on screenshots for GUI task automation.
We found a key challenge in developing such visual GUI agents: GUI grounding - the capacity to accurately locate screen elements based on human instructions.
To address this challenge, we propose to enhance \seek via GUI grounding pre-training, and devise methods to automate the curation of GUI grounding data from web and mobile.
For benchmarking the progress in GUI grounding, we created \seesp, the first realistic evaluation dataset encompassing mobile, desktop, and web platforms.
Results on \seesp demonstrate a significant improvement of \seek over LVLM baselines.
Moreover, comprehensive evaluations across three GUI automation tasks consistently support our finding that advancements in GUI grounding directly correlated with improved performance in downstream agent tasks.

\section*{Limitations}
\seek currently simplifies the GUI action space to mainly focus on clicking and typing, excluding complex actions like dragging and double-clicking. 
Additionally, limited by the performance of open-source LVLMs, training on agent-specific data is necessary for \seek to execute multi-step tasks on interfaces like mobile and computer.

\section*{Ethical considerations}
GUI agents are developed to automate tasks and enhance efficiency on digital devices. These technologies are especially significant for individuals with visual impairments. 
Here are some ethical considerations:

\noindent
\textbf{Privacy Issues.} The operation of GUI agents involves accessing and interacting with user interfaces that may contain personal or sensitive information. Ensuring data protection and user consent are paramount to maintaining privacy integrity.

\noindent
\textbf{Safety in Read-World Interactions.} When GUI agents interact with the real world, there's a risk of unintended harmful actions. Ensuring these agents operate within safe parameters is crucial to prevent negative outcomes.

\noindent
\textbf{Bias.} The development of GUI agents must address potential biases in their algorithms, which could result in unequal performance across different user groups or interface designs. Mitigating bias is essential for equitable access and effectiveness.

Addressing these concerns requires ongoing research and development efforts, ensuring that the benefits of GUI agents are realized without compromising ethical standards.

\clearpage
\bibliography{anthology,custom}
\bibliographystyle{acl_natbib}

\newpage
\appendix

\section{Details of \seek Pre-training}
\label{sec:pretraining}

\subsection{Pre-training Tasks}
\seek employs pre-training tasks as outlined in \Cref{tab:tasks}. For the grounding task, we incorporate two forms: predicting center point coordinates (text\_2\_point) and predicting bounding box (text\_2\_bbox). For the task of generating text for elements (similar to OCR), we also include two categories: predicting text based on center point (point\_2\_text, widget captioning) coordinates and based on bounding boxes (bbox\_2\_text). 
Our preliminary experiments indicated that predicting points was slightly better than bounding boxes, likely due to the variable sizes of UI elements. Consequently, we increased the proportion of data with point localization.
Finally, about 1 million samples are used for the continual pre-training of \seek.

\begin{figure}[t!]
\centering
\includegraphics[width=0.485\textwidth]{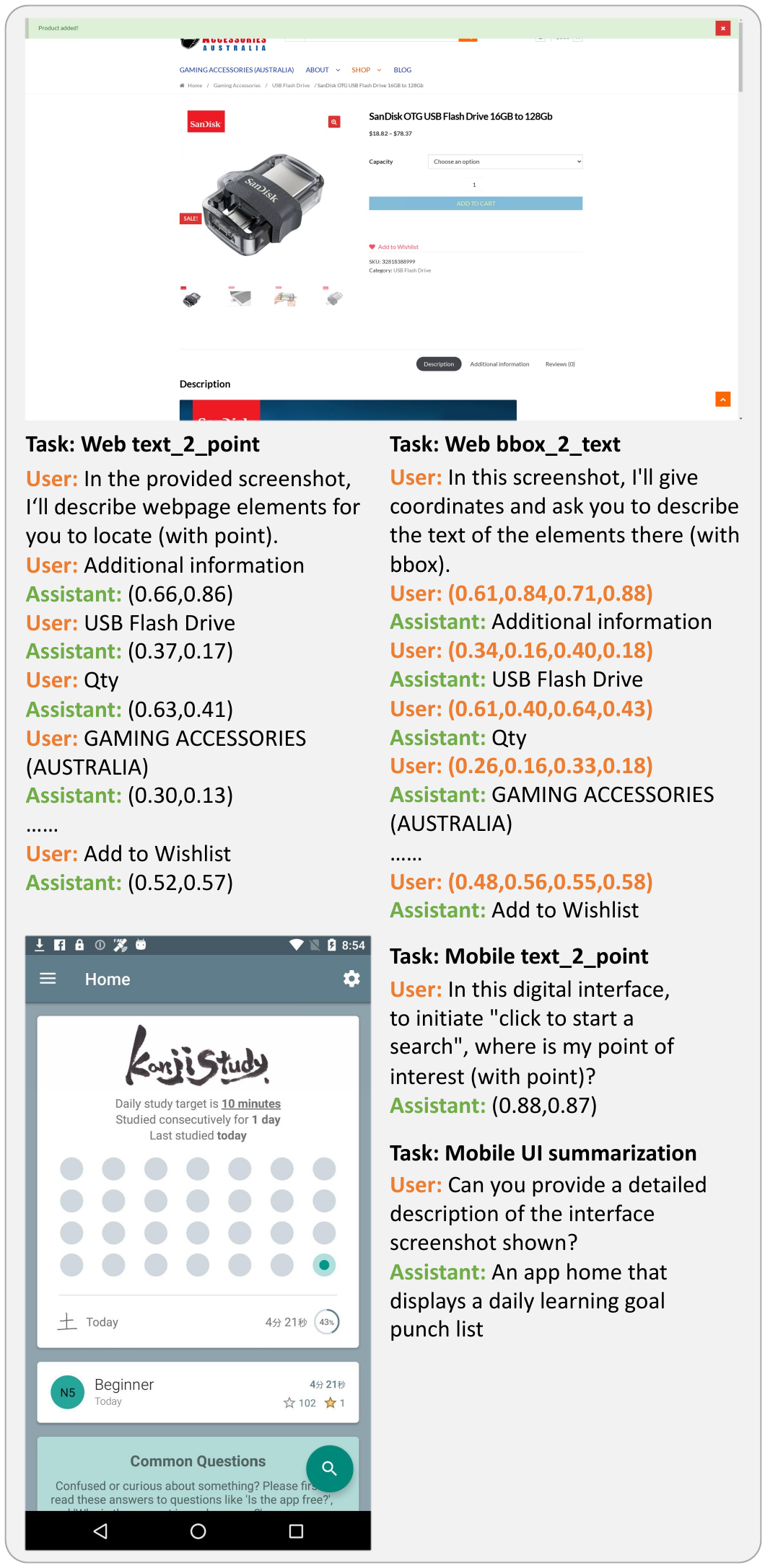}
\caption{Examples of \seek pre-training tasks.}
\label{fig:pretrain_task}
\end{figure}

For tasks involving coordinates, positions are represented as either the point (x,y) or the bounding box of (left, top, right, down), where each value is a two-decimal place number in the range [0,1] indicating the ratio of the corresponding position to the width or height of the image. \Cref{fig:pretrain_task} provides some examples of the pre-training data.

\begin{table}[h]
\centering
\renewcommand\arraystretch{1.2}
\tabcolsep=0.06cm
\begin{tabular}{ccc}
\hline
Domain                  & Task              & Sample Num \\ \hline
\multirow{4}{*}{Web}    & text\_2\_point    & 271K       \\
                        & text\_2\_bbox     & 54K        \\
                        & point\_2\_text    & 54K        \\
                        & bbox\_2\_text     & 54K        \\ \hline
\multirow{4}{*}{Mobile} & text\_2\_point    & 274K        \\
                        & text\_2\_bbox     & 56K        \\
                        & UI summarization  & 48K        \\
                        & widget captioning & 42K        \\ \hline
General                 & LLaVA             & 145K       \\ \hline
\multicolumn{2}{c}{Total}                   & 1M         \\ \hline
\end{tabular}
\caption{All training data used by \seek.}
\label{tab:tasks}
\end{table}

\vspace{-0.30em}
\subsection{Training Configurations}
We employed the aforementioned data for continual pre-training of Qwen-VL-Chat to develop \seek. To enhance LVLM's understanding of GUI images, we unlocked the gradients of its visual encoder and applied LoRA for fine-tuning.
We adopt AdamW as the optimizer and use a cosine annealing scheduler with an init learning rate of 3e-5 and a global batch size of 64. All training takes around 24 hours on 8 NVIDIA A100 GPUs.

\begin{figure*}[t!]
\centering
\includegraphics[width=\textwidth]{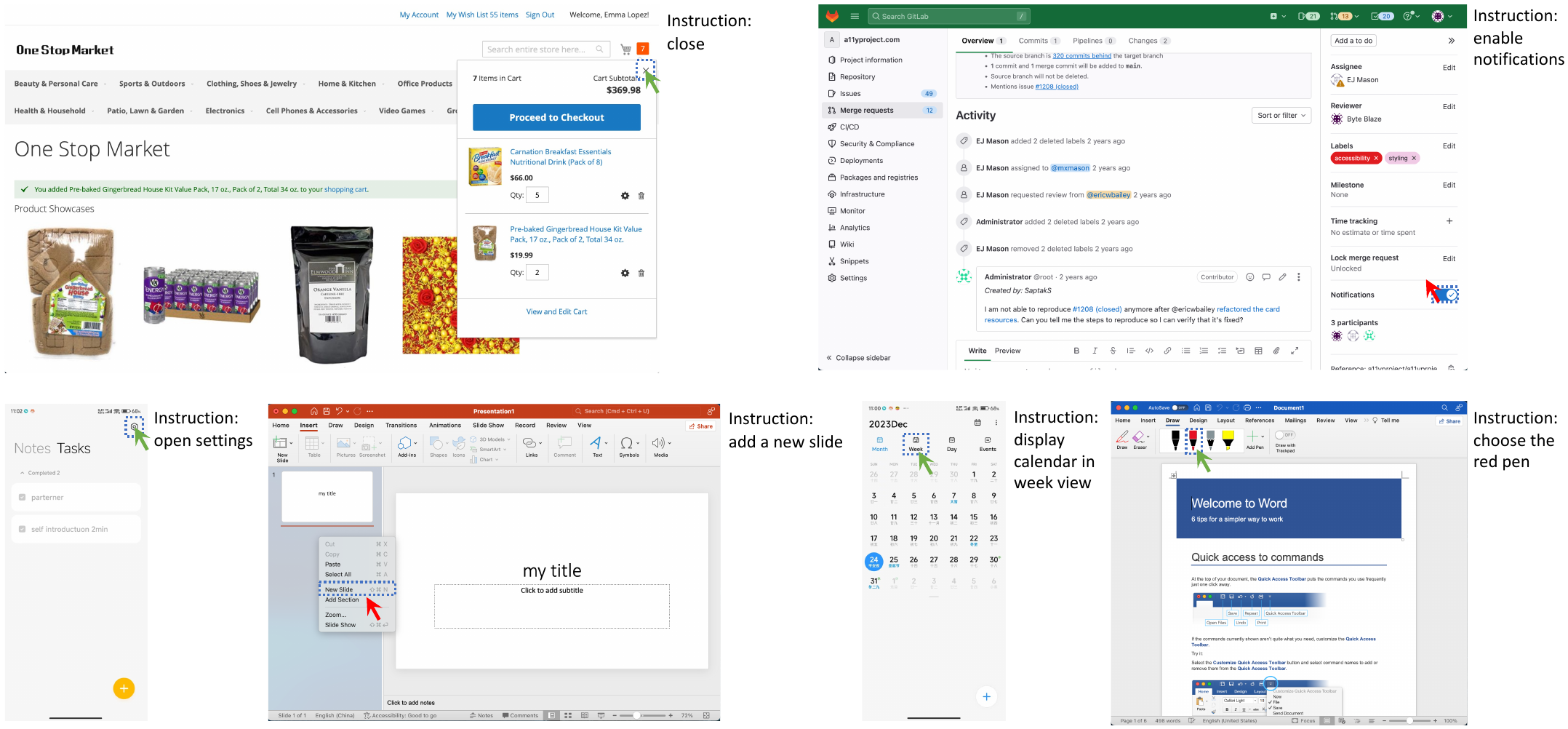}
\caption{\seek on \seesp. Blue dashed boxes represent the ground truth bounding boxes, while green and red pointers indicate correct and incorrect predictions.}
\label{fig:screenspot_seeclick_case}
\vspace{-1.0em}
\end{figure*}

\section{\seesp Annotation \& Evaluation}
\label{app:screenspot}

\subsection{Human Annotation}
We convened four experienced annotators, all either Ph.D. or master students in computer science, proficient in using mobile phones and computers and familiar with GUI operations.
Initially, we assigned different GUI types to the annotators, such as iOS, Windows, and Web.
Then, annotators were required to capture screenshots during their routine use (e.g., various apps) and subsequently annotate the clickable regions of frequently interacted elements using bounding boxes with annotation tool \footnote{http://makesense.bimant.com}. 
Finally, these annotators were instructed to write corresponding English text commands for the annotated screen elements.
All annotated interfaces and operational elements were in English and post-processed to remove personal information.

\subsection{Sample Showcase}
\begin{figure}[t!]
\centering
\includegraphics[width=0.95\textwidth]{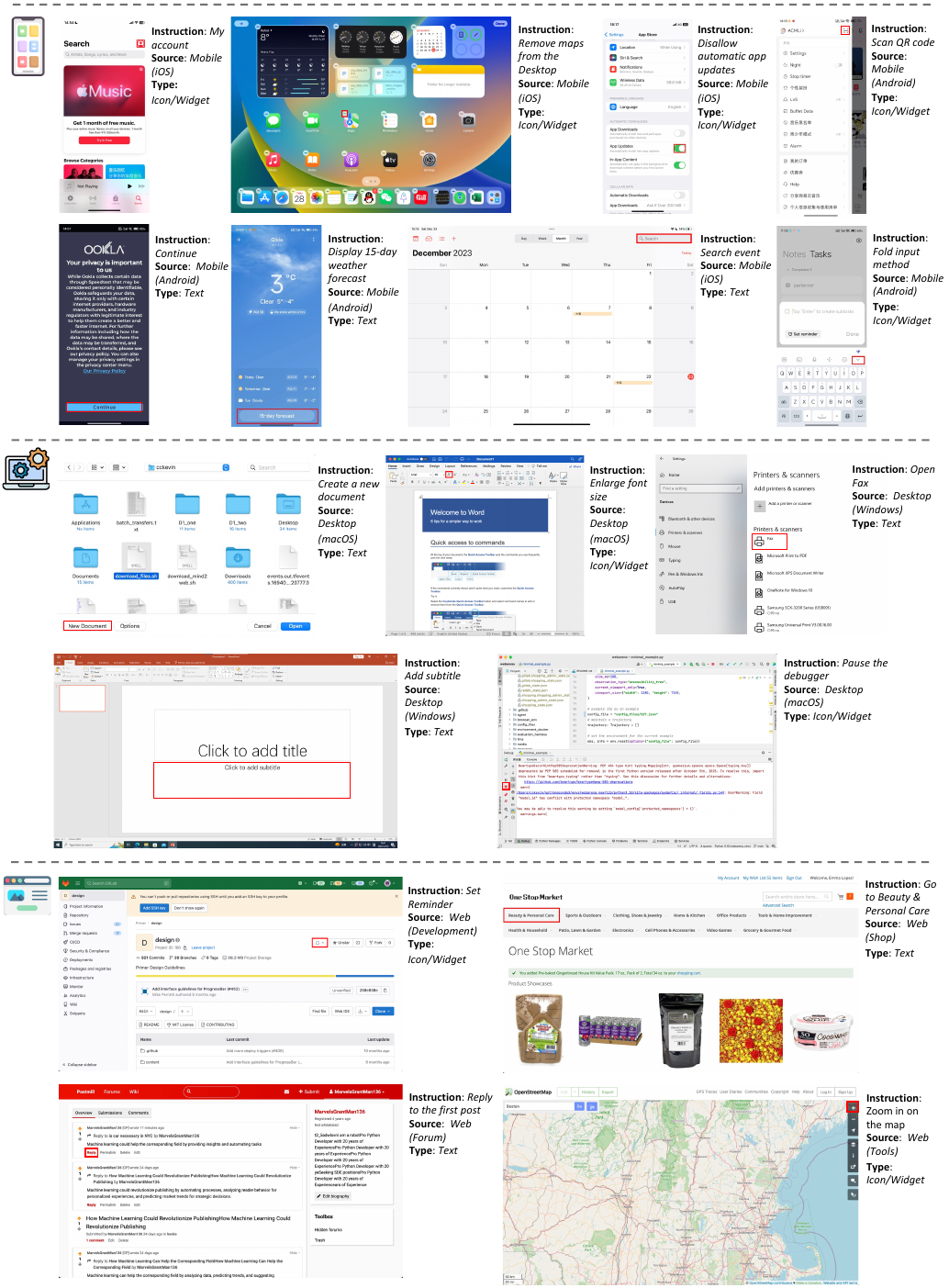}
\captionsetup{justification=centering, singlelinecheck=off}
\begin{minipage}{\textwidth}
    \caption{More examples of GUI grounding benchmark \seesp.}
    \label{screenspot_case}
\end{minipage}
\end{figure}

\Cref{screenspot_case} provides more examples of \seesp, which contains a variety of common GUI scenarios for mobile, desktop, and web platforms.

\subsection{Evaluation Detail}
For comparing baselines, we tested the models' grounding capabilities using their officially recommended approach. For instance, with CogAgent, we randomly selected prompts from the official set provided, such as "What steps do I need to take to <instruction>? (with grounding)", then the output coordinates (or the centers of bounding boxes) were taken as predicted points.
For GPT-4V, we follow \citet{yang2023dawn} to enable it to locate screen elements based on instructions.
\seek's predictions with points were marginally better than bounding boxes, thus we selected point prediction for final evaluation.

\subsection{\seek Case Study \& Error Analysis}
\begin{figure}[t!]
\centering
\includegraphics[width=0.40\textwidth]{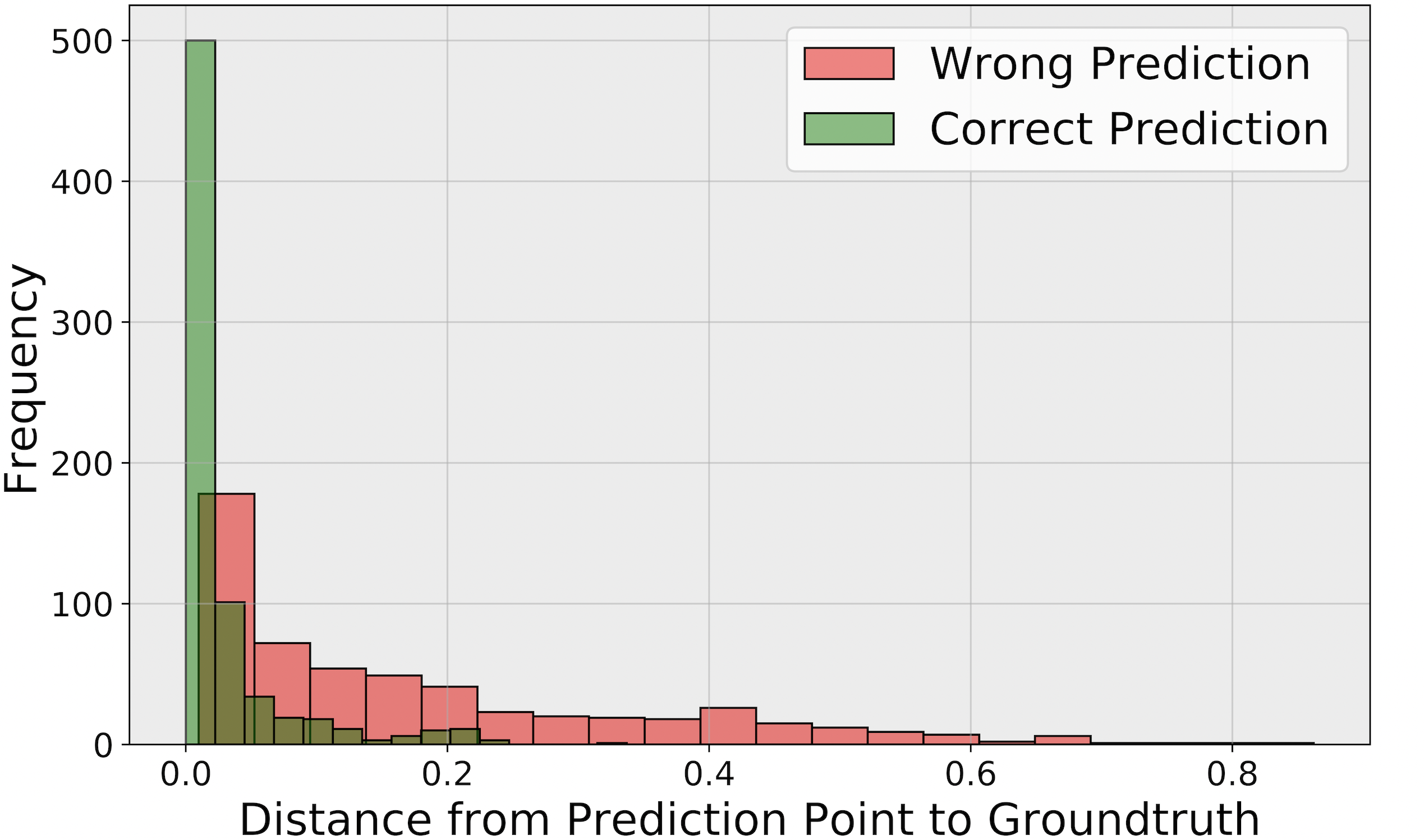}
\caption{Distance distribution of prediction point to ground truth. 
Most incorrect predictions are also close to the answer, suggesting the model recognizes the target but needs improvement in fine-grained localization.
}
\label{fig:screenspot_analyse}
 \vspace{-.35cm}
\end{figure}

\Cref{fig:screenspot_seeclick_case} presents some examples of \seek on \seesp. \seek can comprehend human instructions and accurately locate screen elements.
To conduct a detailed analysis of localization performance, we quantified the distances between predicted points and ground truth (the center of target elements) in \Cref{fig:screenspot_analyse}. 
It's noteworthy that even incorrect predictions mostly occur near the target bounding box, suggesting the model recognizes the target but needs improvement in fine-grained localization.

\section{Downstream Agent Tasks}
\label{app:agent_tasks}
In this section, we first detail the formulation of \seek as a visual GUI agent, then separately introduce the settings for three downstream tasks, and finally show \seek's interaction cases with the GUI across these tasks.

\vspace{-0.25em}
\subsection{Formulation of \seek as Visual GUI Agent}

\noindent
\textbf{Action Space}
\seek involves common human-UI interaction operations. Following AITW, we assigned an \texttt{action\_type id} to each action type for model prediction. 
\begin{itemize}[itemsep=4pt,topsep=4pt,parsep=0pt]
    \item \texttt{click(x,y)}: \texttt{4}. A click action at (x,y), where each value is a [0,1] number indicating the ratio of the corresponding position to the width or height of the image.
    \item \texttt{type("typed\_text")}: \texttt{3}. An action of typing a piece of text.
    \item \texttt{select("value")}: \texttt{2}. An action for selecting an option from a dropdown menu on a webpage.
    \item \texttt{swipe(direction)}: Swipe actions for the screen, swipe up/down/left/right are assigned the ids \texttt{1}, \texttt{0}, \texttt{8}, and \texttt{9} respectively.
    \item \texttt{PRESS BACK}: \texttt{5}. The action for returning to the previous step.
    \item \texttt{PRESS HOME}: \texttt{6}. The action for returning to the homepage.
    \item \texttt{PRESS ENTER}: \texttt{7}. The action of pressing the ENTER key to submit input content.
\end{itemize}
The first two actions, clicking and typing, are universally applicable across various GUIs. The third action, select, is defined according to the specifications in Mind2Web. The latter four actions, along with two additional states, \texttt{TASK COMPLETE} and \texttt{TASK IMPOSSIBLE}, are defined following the AITW framework for Android environments.

\vspace{0.40em}
\noindent
\textbf{Agent Formulation}
\seek is an autonomous agent capable of executing human instructions on GUIs. It takes as input the instruction, a screenshot of the current interface and a series of (k=4 in our setting) previous actions, to predict the next action to be taken. Specifically, \seek uses the following prompt to execute each step of the agent:

\begin{figure}[h!]
\centering
\vspace{-0.50em}
\includegraphics[width=0.50\textwidth]{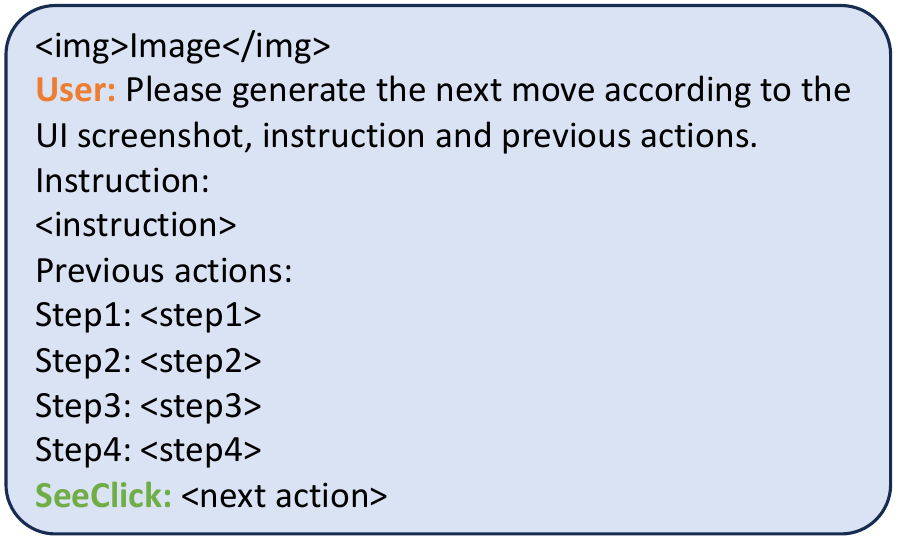}
\vspace{-0.90em}
\label{fig:task_prompt}
 \vspace{-.35cm}
\end{figure}
During training and testing, we organize the data by step into the format described above.

\subsection{MiniWob}
\label{app:miniwob_tasks}
MiniWob is a classic simplified web agent environment, built on Chrome, allowing low-level operations such as clicking and typing.
It comprises around 100 tasks, where each task can templatize random variants and corresponding instructions controlled by a random seed, creating up to billions of possible task instances.
We use 50 successful trajectories for each task provided in \citep{zheng2023synapse} for training and test each task with 50 random seeds, following standard practices.

We report the average success rate across random seeds and tasks, automatically provided by the MiniWob environment.
A task is considered successfully completed if executed correctly, while incorrect executions or exceeding the maximum number of actions (set as 30 here) are counted as failures. 
For the baselines in \Cref{tab:miniwob}, we use the task-wise scores provided in their papers to calculate the average score for tasks overlapping with \seek. We also provided a task-wise comparison in \Cref{tab:55_tasks}.

\begin{table}[t!]
\centering
\renewcommand\arraystretch{1.1}
\setlength\tabcolsep{2pt}
\scriptsize
\resizebox{\columnwidth}{!}{%
\begin{tabular}{ccccccc}
\hline
         & Gen.       & Inst.       & GApps.    & Sing.        & WShop.   & Ovr.       \\ 
         \hline
Auto-UI  & \textbf{68.2} & 76.9          & 71.4          & 84.6          & 70.3          & 74.3          \\
CogAgent & 65.4          & 78.9          & 75.0          & \textbf{93.5} & 71.1          & \textbf{76.9} \\
\seek & 67.6          & \textbf{79.6} & \textbf{75.9} & 84.6          & \textbf{73.1} & 76.2          \\ 
\hline
\end{tabular}
}
\caption{Comparison on the origin split of AITW.}
\label{tab:origin_aitw}
\end{table}

\begin{table*}[t]
\centering
\renewcommand\arraystretch{1.1}
\small
\begin{tabular}{lcccccc}
\hline
                            & CC-Net (SL) & WebN-T5    & WebGUM     & Pix2Act    & Qwen-VL    & \seek   \\ \hline
Choose-date                 & 0.12        & 0.00       & 0.13       & 0.06       & 0.0        & 0.02       \\
Click-button                & 0.78        & 1.0        & 1.0        & 0.32       & 0.42       & 0.96       \\
Click-button-sequence       & 0.47        & 1.0        & 1.0        & 1.0        & 0.08       & 0.86       \\
Click-checkboxes            & 0.32        & 0.96       & 1.0        & 0.99       & 0.44       & 0.78       \\
Click-checkboxes-large      & 0.0         & 0.22       & 0.99       & 1.0        & 0.0        & 0.02       \\
Click-checkboxes-soft       & 0.04        & 0.54       & 0.98       & 0.91       & 0.06       & 0.22       \\
Click-checkboxes-transfer   & 0.36        & 0.63       & 0.99       & 0.76       & 0.60       & 0.70       \\
Click-collapsible-2         & 0.17        & 0.00       & 0.95       & 0.31       & 0.0        & 0.48       \\
Click-collapsible           & 0.81        & 0.00       & 0.98       & 0.80       & 1.0        & 1.0        \\
Click-color                 & 0.82        & 0.27       & 0.34       & 0.88       & 0.96       & 1.0        \\
Click-dialog                & 0.95        & 1.0        & 1.0        & 0.12       & 0.96       & 1.0        \\
Click-dialog-2              & 0.88        & 0.24       & 0.43       & 0.73       & 0.84       & 1.0        \\
Click-link                  & 0.59        & 1.0        & 1.0        & 0.86       & 0.0        & 0.90       \\
Click-option                & 0.21        & 0.37       & 1.0        & 0.0        & 0.70       & 1.0        \\
Click-pie                   & 0.15        & 0.51       & 0.99       & 0.81       & 0.16       & 0.80       \\
Click-shades                & 0.04        & 0.0        & 0.0        & 0.76       & 0.0        & 0.02       \\
Click-shape                 & 0.11        & 0.53       & 0.72       & 0.19       & 0.04       & 0.52       \\
Click-tab                   & 0.95        & 0.74       & 1.0        & 0.54       & 1.0        & 1.0        \\
Click-tab-2                 & 0.27        & 0.18       & 0.95       & 0.52       & 0.0        & 0.60       \\
Click-tab-2-hard            & 0.19        & 0.12       & 0.95       & 0.0        & 0.16       & 0.42       \\
Click-test                  & 1.0         & 1.0        & 1.0        & 1.0        & 1.0        & 1.0        \\
Click-test-2                & 0.95        & 1.0        & 1.0        & 1.0        & 0.72       & 0.94       \\
Click-widget                & 0.56        & 1.0        & 1.0        & 0.87       & 0.38       & 0.58       \\
Count-shape                 & 0.21        & 0.41       & 0.68       & 0.0        & 0.20       & 0.28       \\
Copy-paste                  & 0.04        & -          & -          & -          & 0.96       & 0.80       \\
Copy-paste-2                & 0.01        & -          & -          & -          & 0.96       & 0.80       \\
Email-inbox                 & 0.09        & 0.38       & 0.99       & -          & 0.08       & 0.80       \\
Email-inbox-forward-nl      & 0.0         & 0.6        & 1.0        & -          & 0.24       & 0.74       \\
Email-inbox-forward-nl-turk & 0.0         & 0.33       & 1.0        & -          & 0.16       & 0.56       \\
Email-inbox-nl-turk         & 0.05        & 0.23       & 0.98       & -          & 0.40       & 0.68       \\
Enter-date                  & 0.02        & 0.0        & 1.0        & 0.59       & 1.0        & 1.0        \\
Enter-password              & 0.02        & 0.97       & 1.0        & -          & 1.0        & 1.0        \\
Enter-text                  & 0.35        & 0.89       & 1.0        & -          & 1.0        & 1.0        \\
Enter-text-dynamic          & 0.39        & 0.98       & 1.0        & -          & 0.96       & 1.0        \\
Focus-text                  & 0.99        & 1.0        & 1.0        & -          & 1.0        & 1.0        \\
Focus-text-2                & 0.96        & 1.0        & 1.0        & -          & 0.84       & 0.96       \\
Find-word                   & 0.05        & -          & -          & -          & 1.0        & 0.10       \\
Grid-coordinate             & 0.66        & 0.49       & 1.0        & 0.97       & 0.96       & 0.52       \\
Guess-number                & 0.21        & 0.0        & 0.11       & -          & 1.0        & 1.0        \\
Login-user                  & 0.0         & 0.82       & 1.0        & -          & 1.0        & 1.0        \\
Login-user-popup            & 0.02        & 0.72       & 0.99       & -          & 0.86       & 0.98       \\
Multi-layouts               & 0.00        & 0.83       & 1.0        & -          & 0.44       & 0.72       \\
Multi-orderings             & 0.0         & 0.88       & 1.0        & -          & 0.42       & 0.86       \\
Identify-shape              & 0.68        & -          & -          & 0.94       & 1.0        & 0.68       \\
Navigate-tree               & 0.32        & 0.91       & 1.0        & 0.07       & 0.60       & 0.82       \\
Search-engine               & 0.15        & 0.34       & 0.96       & -          & 0.56       & 0.84       \\
Simple-algebra              & 0.03        & -          & -          & 0.99       & 0.48       & 0.38       \\
Simple-arithmetic           & 0.38        & -          & -          & 0.67       & 0.92       & 0.78       \\
Text-transform              & 0.19        & -          & -          & 0.91       & 0.36       & 0.46       \\
Tic-tac-toe                 & 0.32        & 0.48       & 0.56       & 0.76       & 0.30       & 0.58       \\
Unicode-test                & 0.86        &            &            & 0.64       & 0.54       & 0.98       \\
Use-autocomplete            & 0.07        & 0.22       & 0.98       & 0.95       & 0.72       & 0.82       \\
Use-slider                  & 0.18        & -          & -          & 0.69       & 0.38       & 0.32       \\
Use-spinner                 & 0.47        & 0.07       & 0.11       & -          & 0.24       & 0.16       \\
Read-table                  & 0.01        & -          & -          & -          & 0.90       & 0.72       \\ \hline
Average                     & 0.336 (55)  & 0.552 (45) & 0.861 (45) & 0.646 (35) & 0.564 (55) & 0.712 (55) \\ \hline
\end{tabular}
\caption{Mean scores across 55 MiniWob tasks.}
\label{tab:55_tasks}
\end{table*}

\subsection{AITW}
\label{sec:aitw_origin}

AITW is a recently collected dataset for Android smartphone automation, where each sample comprises an instruction and an action trajectory with screenshots. AITW is divided into five subsets: General, Install, GoogleApps, Single, and WebShopping, totally including over 30K instructions and 700K episodes.

Despite AITW's large scale, as stated in \Cref{sec:aitw_split}, the current train-test split poses a significant risk of overfitting, leading to experimental results that do not accurately reflect an agent's generalization ability in the real world. 
We also conducted experiments on \seek using the origin split, as shown in \Cref{tab:origin_aitw}, \seek is comparable to CogAgent's performance. We believe that due to the severe overfitting, designing new agent frameworks or enlarging model size is unlikely to yield much improvements on this split.

To address the aforementioned issue, we propose to divide the train/val/test in an instruction-wise manner. Specifically, we selected 545/688/306/700/700 instructions from the General/Install/GoogleApps/Single/WebShopping subsets, and retained only one annotated episode for each instruction. 
To avoid imbalance in joint training, we randomly chose 700 instructions from Single and WebShopping. Given the similarity among instructions within Single and WebShopping, these 700 instructions are representative of performance on these two subsets.
Next, we allocate 80\% for training and the remaining 20\% for testing, and select additional 5*100 episodes to form the validation set from the origin data.
The data used for training, validation, and testing will be open-sourced to serve as an effective evaluation.

The other settings are consistent with previous work, calculating a screen-wise matching score that considers both the correctness of the action type and its value (e.g., the click point or typed text). The screen-wise matching score is correlates with the task completion score judged by humans \citep{rawles2023android}.

\subsection{Mind2web}
Mind2Web is a recently proposed dataset for developing generalist web agents for real-world websites, originally designed for text-based agents. Therefore, the origin observation in each step only includes the HTML code of the current webpage.
To train and evaluate visual-based agents, we extracted web screenshots and the bounding boxes of target operational elements for each step from Mind2Web's raw dump.
One issue with Mind2Web's original HTML observation is that it captures the entire page, including scrolling, with its screenshots being long captures (e.g., 1920*12000). 
Predicting operational positions from such high-resolution long screenshots is impractical for current LVLMs and does not align with human operations. 
To address this, for target elements not at the top, we randomly crop around their location, maintaining a consistent screenshot resolution of 1920*1080 for all observed interfaces.

Mind2Web first calculates Element Accuracy (Ele.Acc) which compares the predicted element with groundtruth, and Operation F1 (Op.F1) which calculates the token-level F1 score for the predicted operation. Operation F1 is equivalent to the accuracy of click operations but takes into account the correctness of input values for type and select operations.
For our vision-based approach, Element Accuracy is computed as the accuracy of predicted click points falling in the groundtruth elements' bounding box.
Then, a step is considered successful (Step SR) if both the predicted element and operation are correct.

\subsection{Case Study}
\label{app:casestudy}

\begin{figure}[t!]
\centering
\includegraphics[width=0.95\textwidth]{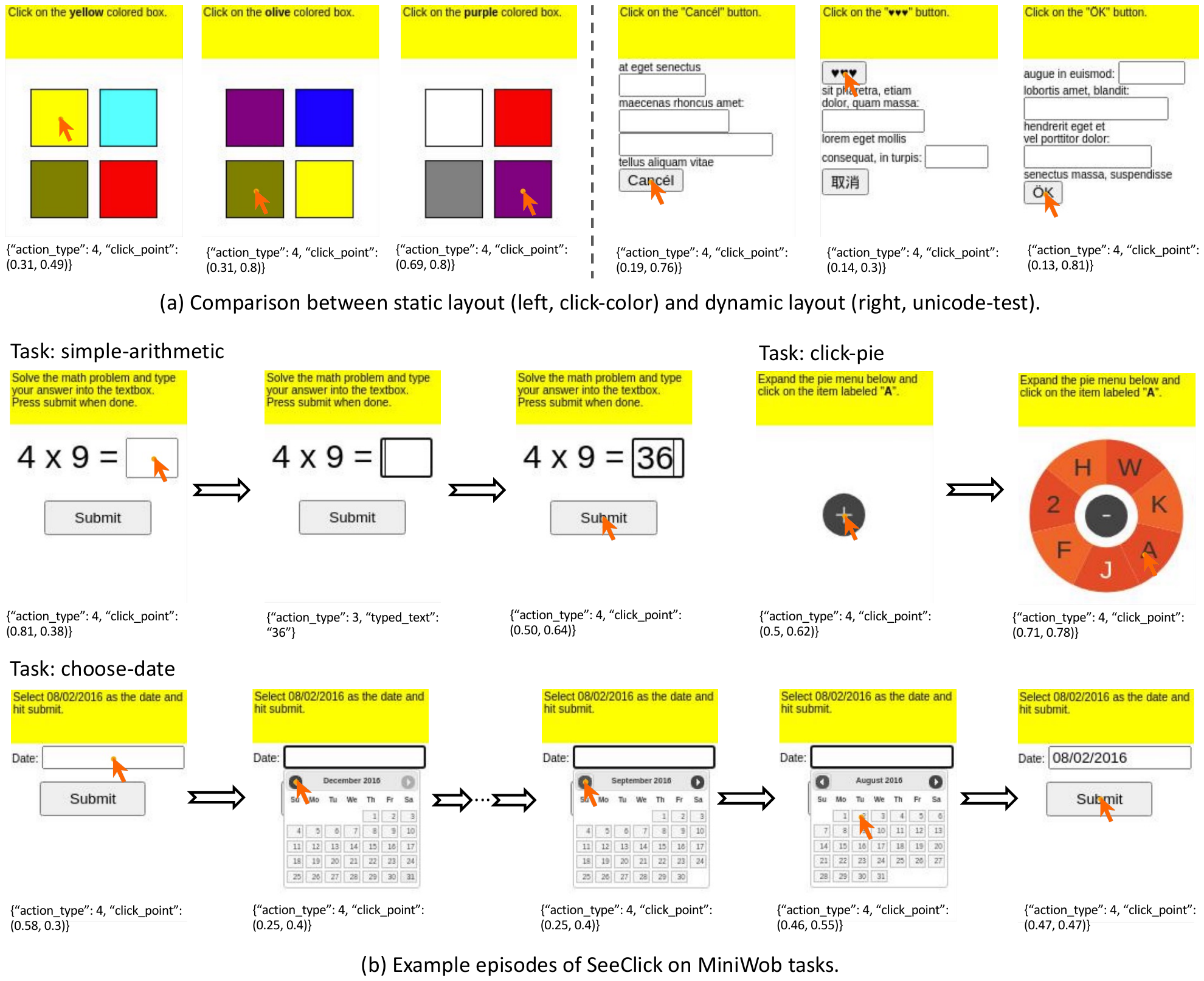}
\captionsetup{singlelinecheck=off}
\begin{minipage}{\textwidth}
    \caption{Example episodes of \seek on MiniWob. The model's prediction output is below the screenshot, with action\_type 4 indicating a click and action\_type 3 indicating typing.}
    \label{miniwob_case}
\end{minipage}
\end{figure}

\noindent
\textbf{MiniWob}
\label{app:miniwob}
\Cref{miniwob_case}(a) illustrates the difference between static and dynamic layout tasks.
Static layout tasks have fixed element positions during training and testing, while dynamic layout tasks display varying interfaces and element positions with instructions, further challenging the agent's ability to accurately locate the target.
\Cref{miniwob_case}(b) provides examples of \seek's interaction with MiniWob. \seek relies solely on the interface screenshot for arithmetic, reasoning, etc.

\begin{figure}[t!]
\centering
\includegraphics[width=0.95\textwidth]{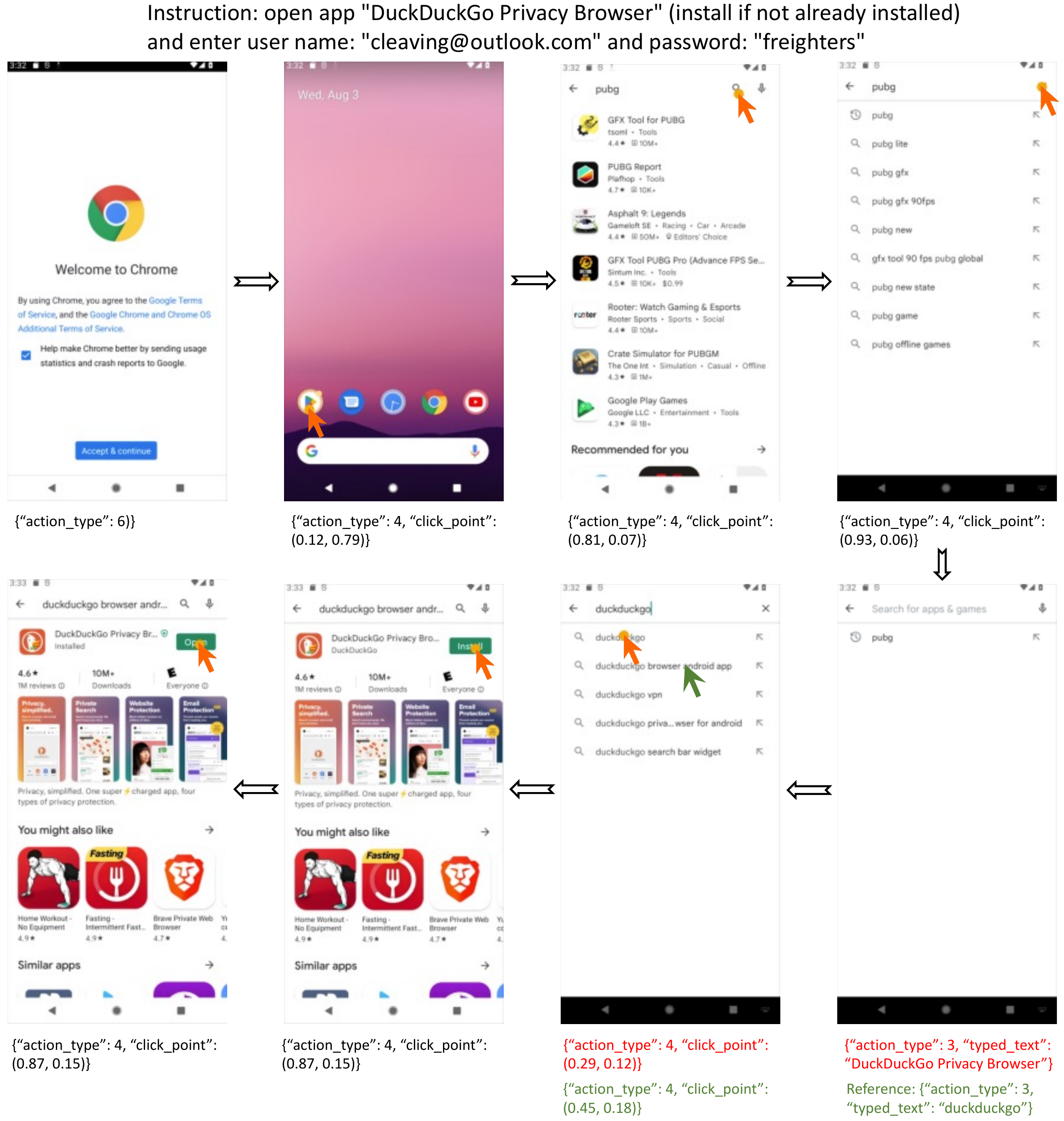}
\captionsetup{singlelinecheck=off}
\begin{minipage}{\textwidth}
    \caption{Example episodes of \seek on AITW. The model's prediction output is below the screenshot, with action\_type 4 indicating a click, action\_type 3 indicating typing and action\_type 6 indicating PRESS HOME. Steps with the red prediction and green reference indicate a failed step.}
    \label{aitw_case}
\end{minipage}
\end{figure}

\noindent
\textbf{AITW}
\Cref{aitw_case} provides \seek's operations on AITW.
Predictions marked in red below indicate that they were computed as incorrect in AITW.
Some errors occur because the current step's answer is not unique.
For example in step 5, the model's predicted input "DuckDuckGo Privacy Browser" is also a potentially correct action.

\begin{figure}[t!]
\centering
\includegraphics[width=0.95\textwidth]{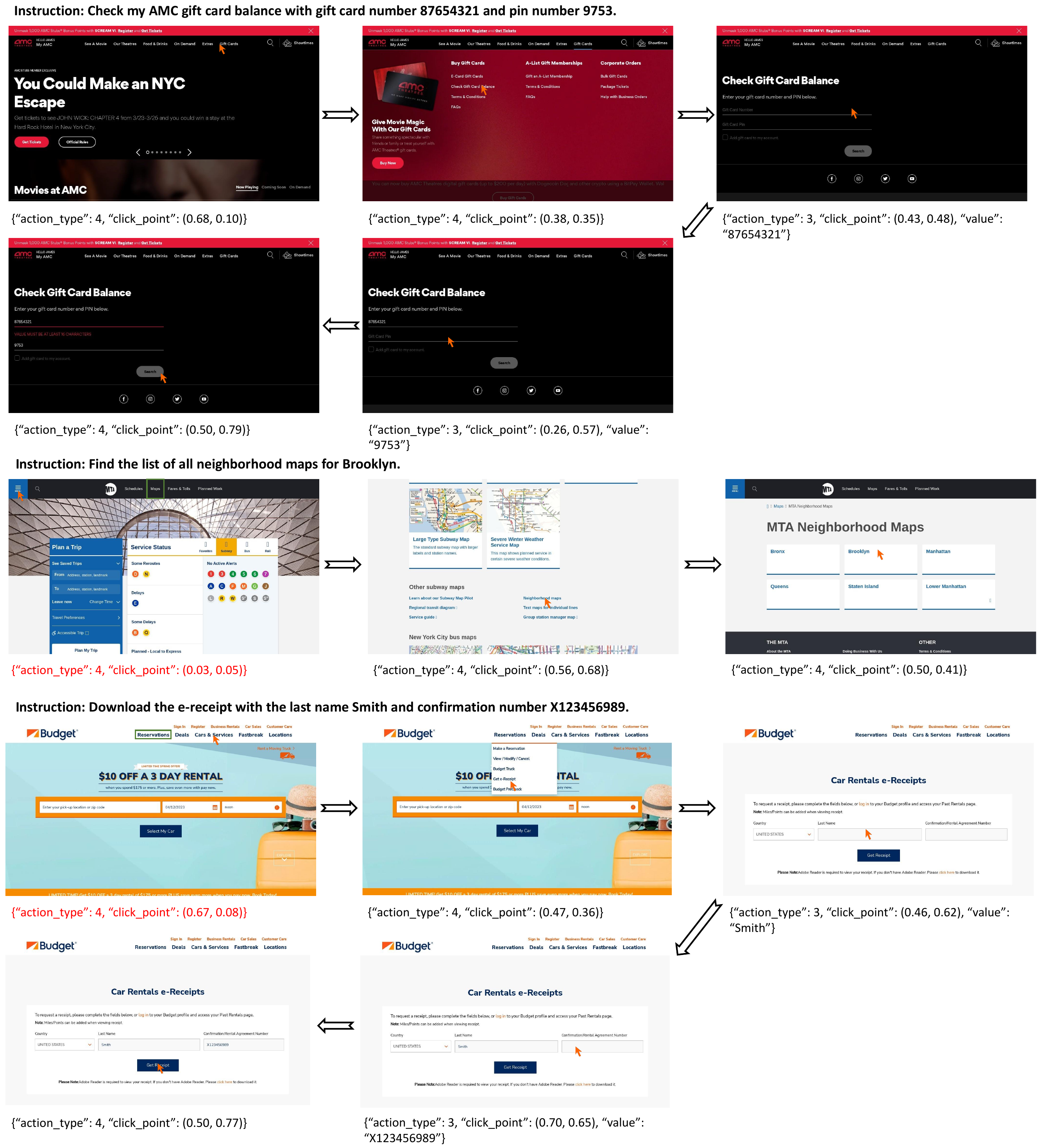}
\captionsetup{singlelinecheck=off}
\begin{minipage}{\textwidth}
    \caption{Example episodes of \seek on Mind2Web. The model's prediction output is below the screenshot, with action\_type 4 indicating a click and action\_type 3 indicating typing. Steps with the red prediction and green reference bounding box indicate a failed step.}
    \label{mind2web_case}
\end{minipage}
\end{figure}

\noindent
\textbf{Mind2Web}
\Cref{mind2web_case} shows several examples of \seek on the real-world website benchmark Mind2Web.
\seek can comprehend instructions and click on the correct elements within complex interfaces.

\clearpage

\clearpage

\clearpage

\clearpage

\end{document}